# Fostering cultural change in research through innovative knowledge sharing, evaluation, and community engagement strategies

*Junsuk Rho[1,2,3], Jinn-Kong Sheu[4], Andrew Forbes[5], Din Ping Tsai[6], Andrea Alú[7,8,9], Wei Li[10,11], Mark Brongersma[12], Joonhee Choi[13], F. Javier Garcia de Abajo[14,15], Laura Na Liu[16,17], Alexander Szameit[18], Tracy Schloemer[19], Andreas Tittl[20,42], Mario Chemnitz[21,22], Cheng Wang[23], Jiejun Zhang[24], Yuri Kivshar[25], Tie Jun Cui[26], Ren-Min Ma[27], Cheng-Wei Qiu[28], Cuicui Lu[29], Yao-Wei Huang[30], Miguel Angel Solis Prosser[31], Ileana-Cristina Benea-Chelmus[32], Rachel Grange[33], Sungjin Kim[34], Anderson S.L. Gomes[35], Davide Ramaccia[36], Yating Wan[37], Apostolos Argyris[38], Antonio G. Souza Filho[39], Tanmoy Chakraborty[40,41], Cristiano Matricardi[32*]*

*Corresponding author: *Cristiano Matricardi, email: cristiano.matricardi@nature.com*

1) *Department of Mechanical Engineering, Pohang University of Science and Technology (POSTECH), Pohang 37673, Republic of Korea*
2) *Department of Electrical Engineering, Pohang University of Science and Technology (POSTECH), Pohang 37673, Republic of Korea*
3) *Department of Chemical Engineering, Pohang University of Science and Technology (POSTECH), Pohang 37673, Republic of Korea*
4) *Department of Photonics, National Cheng Kung University, Taiwan*
5) *School of Physics, University of the Witwatersrand, Johannesburg 2050, South Africa*
6) *Department of Electrical Engineering, City University of Hong Kong, Hong Kong, CHINA*
7) *Electrical Engineering Department, The City College of New York (USA), New York, NY, 10031, USA*
8) *Physics Program, Graduate Center of the City University of New York, New York, NY, 10016, USA*
9) *Photonics Initiative, Advanced Science Research Center, City University of New York, New York, NY, 10031, USA*
10) *GPL Photonics Laboratory, State Key Laboratory of Luminescence Science and Technology, Changchun Institute of Optics, Fine Mechanics and Physics, Chinese Academy of Sciences, Changchun, Jilin, 130033, China.*
11) *University of Chinese Academy of Sciences, Beijing, 100049, China.*
12) *Geballe Laboratory for Advanced Materials, Stanford University, Stanford, California 94305, USA*
13) *Department of Electrical Engineering, Stanford University, Stanford, CA, USA*
14) *ICFO-Institut de Ciencies Fotoniques, The Barcelona Institute of Science and Technology, 08860, Castelldefels, Barcelona, Spain*
15) *ICREA-Institució Catalana de Recerca i Estudis Avançats, Passeig Lluís Companys 23, 08010, Barcelona, Spain*
16) *Second Physics Institute, University of Stuttgart, Pfaffenwaldring 57, 70569 Stuttgart, Germany*
17) *Max Planck Institute for Solid State Research, Heisenbergstrasse 1, 70569 Stuttgart, Germany*
18) *Institute for Physics, University of Rostock, Albert Einstein Strasse 23, 18059 Rostock, Germany*
19) *Department of Electrical Engineering, Stanford University, Stanford, CA USA 94305*
20) *Institute of Photonics, Hamburg University of Technology, 21073 Hamburg, Germany*
21) *Leibniz-Institute of Photonic Technology, Albert-Einstein-Str. 9, 07745 Jena, Germany*
22) *Institute of Applied Optics and Biophysics, Philosophenweg 7, 07743 Jena, Germany*
23) *Department of Electrical Engineering & State Key Laboratory of Terahertz and Millimeter Waves, City University of Hong Kong, Kowloon, China*
24) *College of Physics & Optoelectronic Engineering, Jinan University, Guangzhou, China*
25) *Research School of Physics, Australian National University, Canberra ACT 2601 Australia*

26) State Key Laboratory of Millimeter Waves and Institute of Electromagnetic Space, School of Information Science and Engineering, Southeast University, Nanjing 210096, China
27) School of Physics, Peking University, Beijing, China
28) Optical Science and Engineering Center, National University of Singapore, Republic of Singapore
29) Key Laboratory of Advanced Optoelectronic Quantum Architecture and Measurements of Ministry of Education, Beijing Key Laboratory of Nanophotonics and Ultrafine Optoelectronic Systems, Center for Interdisciplinary Science of Optical Quantum and NEMS Integration, School of Physics, Beijing Institute of Technology, Beijing 100081, China
30) Department of Photonics, College of Electrical and Computer Engineering, National Yang Ming Chiao Tung University, Hsinchu 300093, Taiwan
31) Departamento de Ciencias Físicas, Facultad de Ingeniería y Ciencias, Universidad de La Frontera, Temuco, Chile.
32) Hybrid Photonics Laboratory, École Polytechnique Fédérale de Lausanne, Lausanne, Switzerland
33) ETH Zurich, Department of Physics, Institute for Quantum Electronics, Optical Nanomaterial Group, Auguste-Piccard-Hof, 1, 8093, Zurich, Switzerland
34) Samsung Science and Technology Foundation, Seul, Korea
35) Departamento de Física, Universidade Federal de Pernambuco, 50670-901, Recife, PE, Brazil
36) Department of Industrial, Electronic and Mechanical Engineering, RomaTre University
37) Computer Electrical and Mathematical Sciences and Engineering, King Abdullah University of Science and Technology, Thuwal 23955, Saudi Arabia
38) Institute for Cross-Disciplinary Physics and Complex Systems - IFISC (CSIC-UIB), Palma de Mallorca, Spain
39) Departamento de Física, Universidade Federal do Ceará, 60455-870, Fortaleza-CE, Brazil
40) Department of Electrical Engineering, Indian Institute of Technology Delhi, New Delhi, India, 110016
41) Yardi School of Artificial Intelligence, Indian Institute of Technology Delhi, New Delhi, India, 110016
42) Nano-Institute Munich, Faculty of Physics, Ludwig-Maximilians-Universtität München, Munich, Germany
43) Content innovation department, Springer Nature, Heidelberger Platz 3, Berlin, Germany

## Abstract

Bringing together researchers, funders, industry partners, and publishers from 14 countries across 5 continents, we advance the debate around open-science, assessment and learning. We introduce an integrative "open knowledge system" framework linking knowledge production, validation, assessment, and reuse into one ecosystem view, and translate it into actionable recommendations for each stakeholder. Shifting focus to modular, machine-readable knowledge objects, these recommendations are intended to help diagnose misaligned incentives and guide reforms that properly value all scientific contributions.

## The Current Research Ecosystem

The contemporary research ecosystem is a complex and interconnected network that drives knowledge creation and innovation. At its core are researchers working within diverse institutions, universities, national laboratories, industries, and think tanks whose efforts often culminate in the production of scholarly articles, patents, or products. For these research outputs to achieve official recognition and credibility, they must undergo rigorous evaluation and validation processes.

Peer review, managed by publishers, is the most widely accepted process for validating research. Ideally, peer review ensures the integrity and quality of scientific literature by subjecting it to expert scrutiny. This rigorous vetting process aims to ensure that only robust, high-quality research is disseminated to researchers and the broader public. The validation provided by peer review also plays a significant role in securing funding through government grants, career progression for academic researchers, and other professional opportunities. Scientists gain further validation through citation networks, with metrics such as the h-index reflecting a researcher's productivity and citation impact. Citations are also factored into journal´s impact factor, an index often associated with quality for a single journal or weighed in relation to a specific field. The adoption of Open Science practices has also improved many aspects of research. Open science is a collection of practices, such as sharing data, methodologies, and open-access publications, that enhance transparency, reproducibility, and accessibility of research[1]. Open science practices have been found to foster greater collaboration, accelerate discovery, and ensure that scientific knowledge benefits a broader audience[2,3]. Finally, funding agencies and government bodies provide the financial support and policy frameworks necessary for sustained open research efforts[4]. Industry partners also contribute significantly, particularly in applied research and innovation, fostering collaborations that translate scientific discoveries into practical applications.

Despite these strengths, the peer review and its associated metrics of evaluation face several challenges. Researchers are pressured to publish frequently, which can compromise quality[5], leading to less rigorous[6], hard-to-replicate research[7,8] and increasing scientific fraud, as evidenced by rising retractions.[9] It also puts enormous stress on faculty in fundraising and students/postdocs in securing a job in academia or industry. Furthermore, the focus on novelty in high impact journals often leads to a focus on narrative that distorts the true impact of findings[10] while the volume of new research produced year by year leads to information overload and missed opportunities for research collaboration[11,12]. Additionally, the cost of publishing in open-access journals, which is not always covered by funding agencies, can be a significant burden, especially for early-career researchers[13,14]. Finally, evaluations that often rely on bibliometrics like impact factor or citation indexes may not reflect true quality, disadvantaging niche fields.

Moreover, the peer review process can be slow and biased[15], often benefiting well-established researchers while disadvantaging early-career scientists. Recently, AI-assisted peer-reviews have emerged, aiming to streamline the process[16]. However, if not properly controlled, AI involvement may compromise the reliability of review reports, as it might lack the nuanced understanding that human reviewers provide[17]. Recently, there have been instances where authors deliberately embedded hidden messages or prompts, often formatted as white text or rendered in an extremely small font, that remain invisible to human readers but can be detected by AI-based reviewers, leading them to issue favourable recommendations[18].

Improving access to research outputs is the first step, but real transformation comes from improved knowledge-sharing systems that affect how scientific knowledge is learned, interpreted, and internalized by researchers, how it shapes collective understanding, and how it informs evaluation, reuse, and rewards. What matters is whether research contributions can be meaningfully engaged with, examined, and built upon across the research lifecycle. In this perspective article, we bring together a global community of researchers, funding institutions, industrial partners, and publishers from 14 different countries across the 5 continents to introduce the concept of an *open knowledge framework* as a way to rethink how research contributions are structured, validated, and reused. We then translate the framework into practical considerations across key stakeholders, highlighting their responsibilities, incentives, and benefits. Finally, we examine how these roles connect across the research value chain and why cultural change only becomes possible when these connections are addressed together. In this context, research culture should not be understood mainly as shared values or formal principles[19]. Instead, it is shaped by what researchers are able and incentivized to do in practice, through the evaluation criteria, workflows, and infrastructures that determine what is visible, rewarded, and career-relevant.

This Perspective is the outcome of a community-driven process: an initial draft was circulated among contributors spanning academia, funding bodies, industry, and publishing, and was refined through several rounds of open discussion and iterative revision over one year. The framework and recommendations presented here therefore reflect the convergent views of this community rather than the position of any single author or institution.

## Cultural backbone: The open-knowledge framework

Efforts to reform the research system have so far focused largely on improving access to research outputs. In the meanwhile, we are continuing to treat the research paper as the central version of record and the primary unit through which research contributions are recorded, shared, and assessed. As a result, progress has concentrated on openness and transparency without changing the underlying model by which knowledge is structured, evaluated, or reused. What is missing is a foundational framework that operates at this underlying level. We therefore introduce an open knowledge framework as an enabling substrate for cultural change allowing research practices, incentives, and evaluation to evolve together.

*What does open knowledge mean?* Open knowledge refers to the full set of research contributions, concepts and claims, the logic that links them, the methods used to test them, and the evidence (including negative or inconclusive results) that supports them, made available in forms that others can understand, verify, combine, and reuse. In other words, it is not just about opening the door to the final narrative (a paper), but about making the underlying knowledge legible and actionable for both human and machines. Achieving this requires more than access: it requires an enabling framework, shared standards, incentives, governance, and interoperable infrastructure, so that knowledge can be shared, validated, evaluated, and reused seamlessly across the entire community, rather than merely stored in repositories. In this sense, open knowledge is not an alternative to open science; it is the conceptual framework that let the culture of open science practices deliver its full value.

Initiatives led by the Center for Open Science[20] and national efforts such as the UK Reproducibility Network[21] have been instrumental in articulating expectations around open and reproducible research. More recent works often discussed under the umbrella of *Open Science 2.0* extends this agenda by proposing modular, continuously updated research records[22], large-scale community research practices[23], and open-source infrastructures for data linkage, analysis, and collaboration[24]. Collectively, these efforts make clear that effective open science requires moving beyond static articles toward structured, machine-actionable research records.

So, a natural question arises: Why do researchers (especially those early in their careers), still feel that the only way to get a proper recognition is to publish in high-impact journals, even though the earliest agreements on reforming research assessments date back to the early 2000? Why are they disincentivized from openly sharing their knowledge? The reasons are manifold. From a technical point of view, full knowledge sharing is extremely complicated and time consuming. Furthermore, there are challenges related to the hyper-competitive environment where we need to recalibrate how we compete in research and which assets hold the most value. So, the core issue is not a lack of understanding of the advantages of open knowledge sharing, but the fact that current reward structures, validation processes, evaluation practices, and infrastructures are not designed to make use of those advantages. As a result, stakeholders largely operate in silos, optimizing local incentives rather than collective progress. Addressing this misalignment requires a cultural shift that redefines dependencies between actors and clarifies how value is created, recognized, and exchanged.

What follows is not an attempt to restate the benefits of open science, but to outline the concrete conditions under which they can be operationalized, introducing open knowledge frameworks and mapping the roles of key stakeholders along the research value chain (Figure 1):

a) **Researcher**: who generate knowledge, share it (or keep it closed), and contribute as authors, reviewers, and community members.

b) **Institutions and evaluators**: (e.g., universities, hiring/promotion committees, assessment bodies), who define criteria for recognition and career advancement. c)

c) **Funders and policymakers**: who shape incentives through funding calls, mandates, and the conditions under which funding is provided.

d) **Publishers**: who curate and validate research outputs, turning knowledge objects into trusted, provenance-rich, integrity-certified resources and guiding interpretation and direction through expert curation.

e) **Infrastructure providers:** who build/operate the systems through which knowledge is made accessible and reusable.

Throughout history, change has occurred when a new system offers more advantages than the previous one. Achieving the goal conceptualized above requires the collaborative engagement of all relevant stakeholders to ensure that new frameworks are practical and widely accepted. The next discussion will revisit the roles and responsibilities of the main stakeholders, highlighting the advantages that, once internalized, will drive cultural change, starting with small communities and expanding to more complex and broader ones.

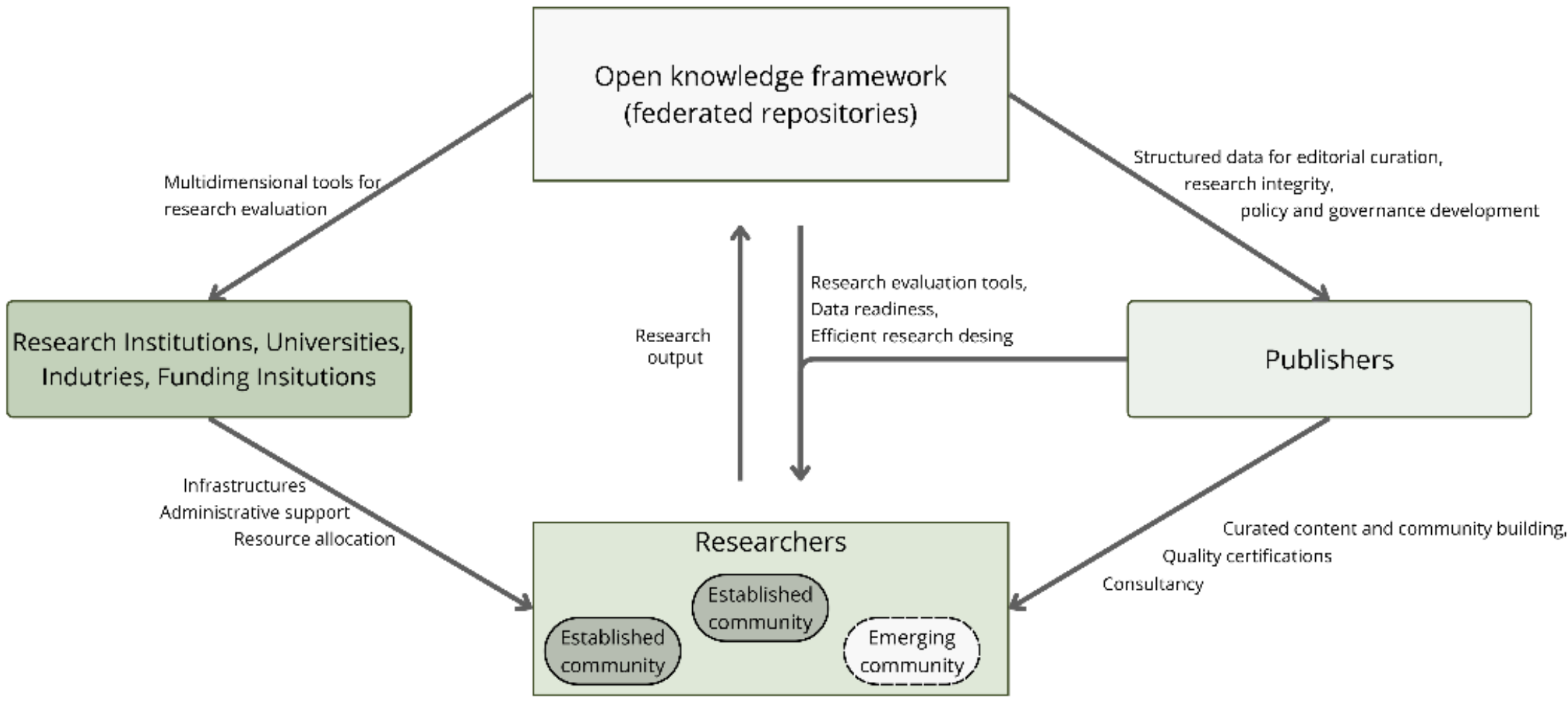

Figure 1. Simplified diagram highlighting the relations between every stakeholder and the conceptualized open knowledge framework. The arrows indicate what a specific group bring to the relative stakeholder.

## The research system

**- Sharing and Recognition** To understand this shift, we must first acknowledge the structural limits of the article and its connected repositories as the primary version of record. PDFs remain effective for reading, but they constrain how research can be validated, reused, evaluated, and acted upon. Data are often separated from figures, figures are locked as static images, and methods, intermediate results, and negative findings are compressed or omitted due to format constraints. These limitations are amplified in an AI-mediated research environment, where missing structure and context increase the risk of misinterpretation and distortion[25]. Rather than replacing the research article, addressing these challenges requires augmenting the version of record with modular, traceable knowledge objects that make core research contributions explicit, operable, and reusable across audiences and tools. Supporting this vision a recent essay from Thibault *et al*. describes a modular and dynamic research record as the fundamental structure of knowledge.[22]
In this framework, it is important to distinguish between open science and open knowledge. Open science refers to a set of practices aimed at making research outputs as accessible as possible, including open access publications, shared data, and open code[1]. However, these practices are largely implemented on top of an article-centric infrastructure that was not designed to represent the full dimensionality of a research work. Open knowledge, by contrast, describes the structural framework that enables open science to function effectively. It is not simply about making more content open, nor should it be confused with sharing compiled manuscripts via preprint servers or repositories. Instead, it treats research not as a single narrative reduction, but as a collection of modular knowledge objects that should be distributed, validated, and reused, as trusted versions of record (or record of version if we talk about open ended knowledge objects and papers). Within such a framework, the article becomes the curated narrative layer that synthesizes and interprets selected contributions, while the underlying knowledge gain value as a reliable, reusable record of the research process (fig.2).

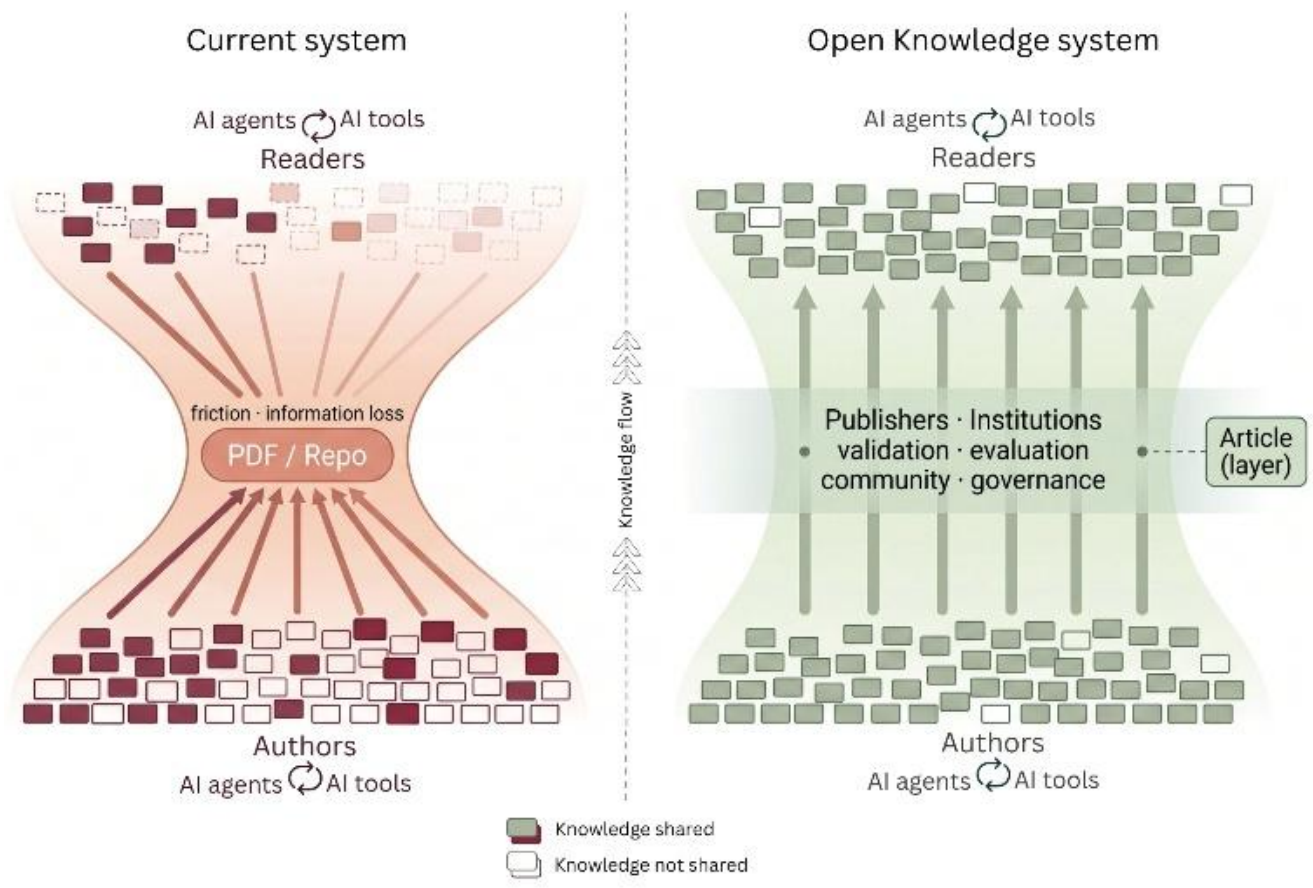

Figure 2. The current and open knowledge systems compared. In the current system (left), most research objects, data, methods, code, negative results, never leave the lab notebook; only a fraction are compressed into a narrative article, creating a bottleneck that generates friction, information loss, and forces readers to reconstruct context. In the open knowledge system (right), structured and interoperable knowledge flows directly from authors to readers and machines through parallel channels, with publishers and institutions acting as a validation and governance layer rather than a gatekeeper. The expansion of shared knowledge objects reflects the core ambition of an open knowledge

framework: making the full dimensionality of research legible, verifiable, and reusable across the entire community.

For this shift to be effective, researchers themselves (not only as authors and readers, but also as editors, peer reviewers, and members of hiring, promotion, and tenure committees) should act as early adopters of structured knowledge sharing and move away from viewing openness primarily through the risk of loss or appropriation. Research communities should establish clear guidelines for early, structured sharing, underpinned by persistent identifiers for both digital (DOIS) and non-digital assets (e.g ORCID for researchers[26], RRIDs for materials[27]). Staged disclosure strategies[28], guided by the principle of "as open as possible, as closed as necessary,"[29] can reduce perceived risks while maximizing opportunities for reuse. In doing so, all contributions gain value for visibility, leadership, and expertise across the research lifecycle. If each disclosed contribution is citable, versioned, and machine-readable, it becomes a unit that the community can reuse, so learning happens earlier and at scale, rather than after the final narrative paper.

These concepts are more concrete than one could think. From the researcher point view the COVID-era collaborations highlight the benefits to science and society of working across borders, cultures and disciplines[30]. Regarding digital infrastructures, a recent example is the European Open Science Cloud, developed in collaboration with OpenAIRE[31], while CoARA´s working groups and national chapters are drawing up the guidelines for evolving researcher evaluation. The digital infrastructure for this change is already being implemented, for example with the European project SciLake[32], a project of the EOSC that aims to build a research ecosystem where knowledge is contextualized, interconnected, interoperable, and accessible. Another relevant example is the NSFC Open Access Repository (OAR), officially launched by the National Natural Science Foundation of China[33], which collects the full texts of research articles resulting from NSFC-funded projects and makes them openly accessible to the public. Similarly, projects like ResearchObjects[34] create platforms to digitize research knowledge and track researchers' work. India is also developing a national open-science ecosystem through coordinated policy and infrastructure[35]. Public funders mandate open access to publicly funded research[36], while national initiatives like "One Nation, One Subscription"[37], and the "National Knowledge Network"[38] expand access to journals, data, and digital research infrastructure, with empirical evidence showing that Indian funding bodies increasingly frame public access as a societal obligation[39]..

India is also building its own ecosystem for open science through both government and private initiatives. National, provide broad access to journals, interoperable datasets, and digital infrastructure across disciplines. Policy mandates are also increasingly explicit: cOAlition S (Plan S)[40] pushes immediate Open Access to funded outputs, while the US OSTP[41] memo directs agencies to ensure free, immediate public access to federally funded research. National funders are also launching complementary initiatives (e.g., Chile's ANID [42], India´s Technology and Innovation Policy[43] ,Open access Korea[44]). Finally, other private initiatives are also bringing innovative solution like DeSci Labs[45] that merge blockchain technology and data sharing for a new research ecosystem while they have also recently launched new evaluation metrics that more fairly evaluate a researcher's contributions. This signals that, the open knowledge framework has the potential to become a cornerstone of the entire research ecosystem, complementing the role of traditional papers for research assessment. However, this is where the involvement of all stakeholders becomes essential. Evaluators, funders, and publishers should contribute to recognizing value through multidimensional evaluation frameworks and differentiated validation processes, rather than through output volume or citation counts alone. This involves reshaping how contributions are assessed and legitimized, an issue explored in more detail in the sections that follow.

**- Virtual Communities**

In today's digital age, virtual communities offer platforms for people to connect, share knowledge, and work together across geographical boundaries. These communities provide spaces where ideas can flourish, diverse perspectives can solve complex problems, and continuous learning and mentorship can thrive. By leveraging virtual communities, we can build inclusive networks that drive progress and create lasting impact (e.g Linux[46]. Wikipedia, Github, etc.).

Furthermore, by integrating open knowledge with virtual communities, we elevate knowledge sharing to a new level of complexity. Virtual communities can enable researchers to share findings, replicate experiments, and discuss results in real-time, potentially accelerating scientific discovery and innovation. By participating, scientists can access a wealth of knowledge and contribute to the collective advancement of their field. This can be extended also to experiments where computer-controlled labs could, in principle, give anyone anywhere in the world time on the experiment [47]. Thus, virtual communities play a crucial role in creating a dynamic and collaborative scientific environment.

## The Evaluation system

Hiring, promotion, and tenure processes play a central role in shaping research culture and represent a critical leverage point for change. In practice, however, they can only reward what evaluation systems are able to observe and validate at scale. Given the limited time, tools, and signals available to evaluators, assessment often relies on a small set of widely recognized proxies, most notably final publications, citation-based indicators, narrative CVs, and recommendation letters[48].

While complementary approaches exist[49,50], the paper-centric model still largely defines what is visible and valued. This reliance on publication- and citation-based indicators reinforces incentives for volume and speed. It can disadvantage interdisciplinary and early-career researchers, as well as contributions whose value emerges over longer time horizons.

Moving beyond these limitations requires evaluation systems that can "see" research at a finer level of granularity. An open knowledge framework, in which research outputs are shared as modular, interlinked knowledge objects within an open infrastructure, would enable this shift. By making research contributions traceable, reusable, and contextualized, this approach supports dependency graphs and knowledge networks and enables a multidimensional evaluation framework (Figure 3b). It allows influence to be measured beyond citations by tracking specific outputs, such as datasets, code, protocols, and validated modules, and their downstream use in research, policy, commercialization, and education. In turn, this creates incentives for researchers to produce and explore contributions that meaningfully support learning across the research community.

This view aligns with a broader understanding of scientific excellence. While being first to make a discovery deserves recognition, primacy alone is not a definitive marker of quality. Scientific reputation is built through reliable, cumulative contributions that others can confidently build upon.

In this regards many institutions are independently working under the Coalition for Advancing Research Assessment (CoARA) to develop strategies for evaluating scientists without heavily relying on publication metrics. The OPUS project for example, aims to advance open-science and research assessment using a stakeholder-driven feedback loop to develop, monitor, refine, and validate evaluations metrics. During this project they have developed a comprehensive evaluation matrix that considers several different aspects of research[50].

While a comprehensive review of current evaluation metrics is outside the scope of this discussion, we will highlight the key factors necessary to foster a change of perspective when we talk about sharing and assessing results as well as their validation through an evolved peer review process

**- Community participation and open-knowledge adoption.**

The central pillar of this system's success lies in fostering communities of researchers to share knowledge (as previously defined) before they become part of a larger narrative. Thus, we should consider several new aspects in the evaluation. In this regard, the frequency and active participation of researchers in the community, as well as the number of research outputs shared, could serve as complementary indicators in research evaluation. This would subject research to community scrutiny before publication, which could enhance the quality and impact of the work, but in particular, opens the door to the next generation of peer review. Indeed, every community will need to specify their own parameters according to how research is normally conducted. But why should researchers spend energy to move to a different system? What are the advantages? Adopting community-based solutions means first and foremost entrusting to community interaction and knowledge sharing with most of the research assessment. This in turn could reduce the urgent need to publish a paper in favour of a system that rewards the actual research work, and the solid knowledge created. Another advantage is the fact that conducting research would be a more organic process where access to all data (positive and negative), ensures the efficient planning of experiments. Furthermore, the possibility to harness the community potential for remote experiments would break geographic and socioeconomic boundaries. Flexibility in this framework should also enable a healthy balance between individual creative thinking and community-based work. A recent example that supports the vision of open knowledge, researcher communities, and recognition of contributions beyond just peer-reviewed papers is the NSERC-funded SiEPIC project led by Prof. Lukas Chrostowski at UBC. The project, involving over ten professors and hundreds of graduate students across Canada, produced open-source outcomes like chip design tools, PDKs, and publications on GitHub, widely used by academic and industrial entities[51].

**- The peer review**

Peer-review remains a cornerstone for validating research findings, yet it must evolve to meet the demands of modern, increasingly complex research landscapes. The surge in the number of researchers and manuscripts, the diversification and convergence of scientific fields, and the emergence of new disciplines have all contributed to a more intricate research ecosystem. Consequently, evidence of inconsistency in peer-review outcomes[52], together with widespread researcher concerns about bias and fairness[53,54], may contribute to perceptions that editorial decisions are unpredictable and may weaken some authors' confidence in the process. Although peer-review has upheld the integrity and advancement of scientific research in the modern era, it is now facing challenges in reliability. The intricate nature of contemporary research can no longer be thoroughly evaluated by a small group of experts. The fast-growing number of papers published each year puts additional strain on the system. The current review process must be segmented and branched out seeking a balance between community-driven, publication-driven and funding-driven research evaluation (Figure 3a). An open-knowledge system could play a mediating role, ensuring that research quality is assessed not only based on citations and journal prestige but also on reproducibility, accessibility, and broader scientific impact and by different stockholders: the community, domain specific experts (e.g. industry) and publishers.

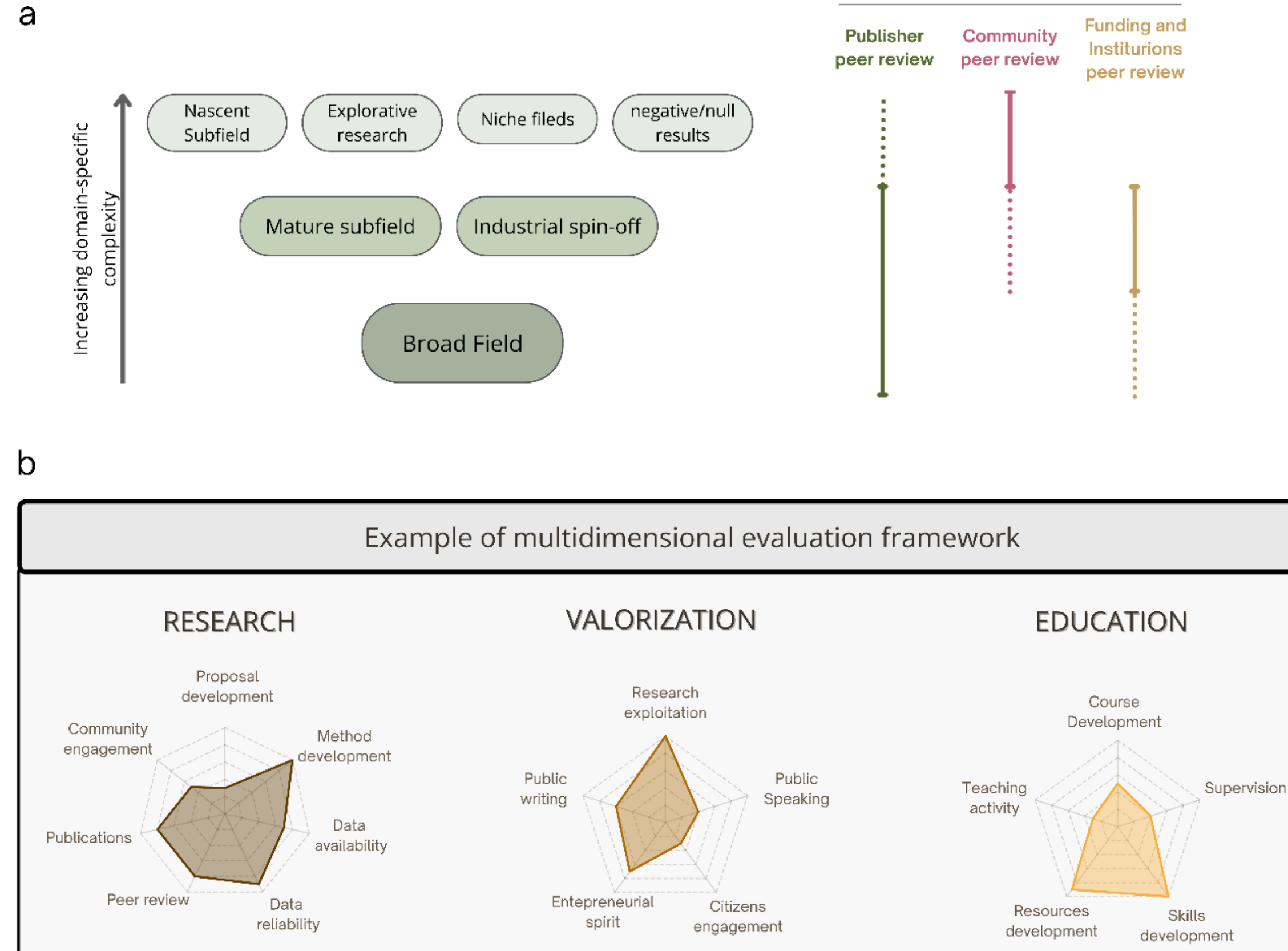

*Figure 3. Schematics illustrations of an evolved per review landscape. a) different domains within the entire scientific ecosystem and the relative adapted peer review strategy, b) outcome of the evaluation process harnessing multidimensional tools for highlighting skills, achievements and attitude. The parameter and evaluation groups have been adapted from the report of the Open and Universal Science Project which identifies indicators and metrics to advance research evaluation[55]*

*i) Community peer review: from being the first to being the best*

Currently, we delegate the task of evaluating both the technical and conceptual aspects of research to peer review processes managed by publishers. Depending on the journal, this process may also assess the novelty, integrity, and potential confirmation biases within the scientific community.

The proposed system relies on a federated knowledge-sharing infrastructure like the one being implemented by EOSC that acts as an open space, offering data, indicators, tools, services, and guidance to support research assessment. Such environments are being developed by projects like SciLake and GRASPOS[56]. In this system, the evaluation will involve many indicators: early and open sharing of research, participation in open peer-review, measures to ensure reproducibility of results, and involving all stakeholders in co-creation, following the criteria outlined by the OPUS project. Following this system, a focus on community peer review can be integrated. This would establish open peer review solutions where researchers can submit results as research modules (rather than full manuscripts), assign a DOI for citation, and subject their findings to community peer review, enabling experts to provide transparent feedback and discussion to assess the scientific validity of the results, before a work is even submitted to a journal. This would foster a reputation system that incentivize thorough and constructive reviews, while ensuring that a diverse pool of reviewers for a well-rounded evaluation. Furthermore, structured review criteria and collaborative review sessions can standardize and enrich the evaluation process like in conferences when showing data to the specific community. Additionally, post-publication review encourages ongoing assessment, adding community feedback to traditional peer review frameworks. This holistic approach will help ensure the free flow of validated knowledge, which is essential for advancing scientific discovery. However, relying solely on community peer review, we risk promoting shady networks and bullying of naturally more isolated researchers, forming sub-communities that act against the broader system. These sub-communities can arise due to diverging interests and goals, leading to echo chambers where only certain views are reinforced. This can stifle innovation and create resistance to new ideas. Additionally, resource allocation might become uneven, with influential sub-communities attracting more resources, causing imbalances and tensions. Sub-communities may also develop conflicting norms and resist broader governance, undermining cohesion. Lastly, the spread of misinformation and misuse of information can harm the community's credibility.

*ii) The role of publishers´ peer review*
To mitigate the above risks, traditional peer review implemented by publisher and certification institutes, although less important for evaluation in open knowledge systems, still plays a vital role in validating research. It provides a structured and rigorous evaluation process that complements open knowledge practices, ensuring that scientific contributions are credible and impactful. By balancing open knowledge-based peer review with professional editorial oversight in traditional peer review, the scientific community can maintain integrity and foster innovation.

*iii) Funding-driven evaluation*
Research communities initially focused on fundamental science may evolve into more application-driven fields due to funding and policy incentives. This transition often leads to increased industry collaboration and, in many cases, the commercialization of knowledge through patents and spin-off companies. While such developments are valuable for technological progress, they also pose challenges to maintaining open-knowledge principles. Researchers in these fields may prioritize patenting over early data sharing to secure competitive advantages, creating a dynamic where early openness is not always feasible. This aspect should also be considered in the evaluation, as creating deep-tech companies or startups based on fundamental research is a great return on investment. These companies must keep a competitive advantage in the form of industrial secrets or patents.

## The publication system

As we outlined earlier, open or community-based evaluation could create new integrity challenges if reviewer identity, provenance, and report quality are not verified[57]. Open dialogue based on early knowledge sharing also could bring the risk of increasing biases between communities by serving the interest of specific groups. However, transparency may also help detect and deter fraud, so the risk should be framed as a governance challenge rather than as evidence that open science itself loses value[58]. For this reason, the need to draft a final manuscript (or any new content type) and publish the work in a peer-reviewed journal will remain. Thus, even though the evaluation will no longer rely solely on published material, the role of scientific publishers will still be central importance for counterbalancing potential community hegemony and independently validating research by adopting the highest peer review standards and adhering to the strictest integrity principles. In this new model, publishers could fully take on the role of content curators and quality certifiers. They would select projects to explore, ensure high-quality standards, create products that help scientists engage with their community, and promote research to the general public. Compared to community knowledge sharing, publishers can operate at a broader level, integrating individual contributions into a larger narrative. They could also provide appropriate consultancy services to help researchers communicate their results and reach the right audience. While this is already true for many top-tier journals, including but not restricted to those in the Nature Portfolio, the problem is that right now everything revolves around one main product: the paper. Conversely, in a system where the evaluation of scientists will be based on many different aspects, the diversification of products, introduction of new services, seminars, and consultancy will rapidly gain higher importance for the development of a researcher. In this environment, a publisher that diversifies and harnesses new technologies could become the reference not only for results certification but also the go-to solution to learn from and enhance their research output. This change could be summarized in a system called publishing-as-a-service. In this framework professional editors, experts who dedicate their full time to developing a broader perspective from a specific field to broader topics, will help researchers contextualize each development in the field. Through a rational process editors will assess and make an informed decision not only based on pure scientific development but also defining whether and how that specific development fits into a bigger picture. This task, which is completely different from making a technical evaluation (mainly performed with a community peer review), is complementary to the job of a researcher and will define the identity of a journal.

## The funding system

Research funding catalyzes innovation by granting researchers access to advanced technologies, specialized equipment, and expert collaborations. This support transforms ideas into tangible contributions to human knowledge. Funding agencies drive research, guiding its direction through their project selections. Given their crucial role, it is understandable that these agencies seek resources for optimal assessment. However, this overreliance can introduce biases and lead to less informed resource allocation, reducing overall efficiency. Contrary to this approach, some funding agencies instruct reviewers to assess the broader experience of applicants in terms of supervision, training, engagement with the public and education (EPE), and the generation of knowledge. The number of publications and impact factors are typically redacted from applications, emphasizing a more holistic view of research assessment. This approach underscores the commitment to a fairer and more comprehensive evaluation of scientific contributions, ensuring that innovative ideas and emerging voices receive the attention they deserve. In this framework, relying on an open knowledge environment offers significant financial and societal advantages for both public and private funding agencies. By leveraging evolved peer review systems, funding agencies can access stronger evidence to make well-informed decisions. Open knowledge systems would also lead to more efficient use of resources by avoiding duplication of efforts and maximizing the impact of funding. Additionally, both public and private funders can enhance their reputation and credibility by supporting transparent and open research practices, which are increasingly valued by the scientific community and the public.

Beyond adopting open knowledge principles for funding allocations, funding agencies can act as regulators by recommending a publication threshold to ensure the quality of each output. This approach emphasizes that the number of publications is not the only proxy for productivity because instead, productivity will be assessed through open knowledge sharing and community interactions.

## Conclusion: collaborations to enable change

Changes to the knowledge ecosystem require collective action, because no single actor can drive change alone. Additionally, funders, research institutions, and policymakers can support open knowledge initiatives by adopting alternative evaluation methods and mandating public access of research outputs, thereby enabling more efficient resource allocation. Furthermore, leveraging both existing and emerging platforms to create efficient virtual spaces will foster community building, enhance collaboration among researchers, institutions, and publishers, and ultimately strengthen the research process. Finally, the engagement of the wider public is also essential in shaping the future of academic publishing. Increased public awareness and demand for open knowledge frameworks can create a stronger motivation for change within the research community and encourage researchers to prioritize sharing their work openly.

**Call to action: building the modular record of the future, together**

Moving from paper-centric publishing to an open knowledge system will not happen through a single new platform, mandate, or business model. It will emerge when communities agree on shared, machine-readable building blocks for research (modular knowledge objects), make them openly reusable by default, and align validation, incentives, and governance so that everyone can participate without losing credit or control. We therefore invite researchers, evaluators, funders, publishers, and infrastructure providers to make a concrete start wherever possible: test modular, open, machine-readable versions of record in real workflows, and publish what you learn so the ecosystem can converge on interoperable standards and inclusive governance. Role-specific actions include:

- Researchers can start small pilots to share and reuse research contributions in modular, reusable forms within their communities, and help normalize community validation of those contributions.
- Infrastructure providers: build a federated knowledge commons that intercepts contributions at creation time and encodes them as machine-readable, provenance-rich objects that remain citable, traceable, and reusable across communities.
- Institution and evaluators can update hiring, promotion, and tenure criteria to explicitly credit transparent, reusable contributions and community validation of knowledge objects, not only publication venue.
- Funders and policymakers can align mandates and incentives so that reuse-ready outputs, validation, and stewardship are supported and rewarded as part of doing research.
- Publishers should move upstream from post-hoc validation to enable early, continuous validation of

research contributions. They ensure that knowledge meets high standards of research integrity, machine-readability, and community relevance from the outset. Beyond dissemination, they offer curation, quality certification, and consultancy that help communities grow, interact, and remain vibrant in interpreting and reusing research outputs. Through modular, plug-and-play frameworks for validation, dissemination, and reuse, publishers act as independent guarantors of quality, ensuring contributions are traceable, verifiable, and integrated across communities.

Implementing these recommendations in small, well-defined communities is a necessary first step to establish viability and refine practice. Scaling to larger and more diverse communities will introduce substantial challenges, particularly around interoperability and governance; meeting these challenges requires a sustained, coordinated commitment across the research ecosystem to drive cultural change.

**Governance principle:** the infrastructure that carries the modular record must be trusted, open, and inclusive. Ownership should be federated across stakeholders (community-governed standards; plural, interoperable service providers; transparent sustainability models), with explicit support for participation by smaller labs, institutions, and regions. The open knowledge system succeeds only if it lowers barriers and makes every contribution, technical, editorial, evaluative, or infrastructural, visible, attributable, and reusable.

**Limitations.**

This article is a perspective, not a systematic review or an empirical study; the framework and recommendations reflect the considered consensus of the contributing community, developed through iterative discussion and revision, rather than quantitative evidence. Although the authors span 14 countries, five continents, and multiple stakeholder roles, the group is weighted toward the physical sciences and related fields, and the applicability of our recommendations to other domains (e.g., clinical research, humanities) may differ and deserves dedicated analysis.

The open knowledge framework we propose is conceptual: its feasibility, costs, and potential unintended consequences, such as metrics, added workload for researchers, or the community-fragmentation risks we discuss, have not been empirically tested and will only become clear through the small-scale pilots we advocate. Several questions remain open and merit dedicated studies. What is the right granularity of a "knowledge object"? Can early, machine-readable sharing of the full research record, including negative results, accelerate collective learning, or does it risk amplifying information overload? How can the quality of a modular contribution be assessed outside the narrative context of a paper? How can multidimensional indicators be protected from strategic inflation, for example through many low-value contributions? How can early adopters be supported? Finally, questions of technical standardization and governance are deliberately left open, as we believe they must be resolved by the communities implementing the framework rather than prescribed in advance.

**Acknowledgements**

We would like to thank Dr. Miranda Vinay for their valuable discussions and feedback that greatly improved the work.

**Funding**

This work received no specific grant from any funding agency in the public, commercial, or not-for-profit sectors

**Author contribution**

All authors: J.R., J.-K.S., A.F., D.P.T., A.A., W.L., M.B., J.C., J.G.d.A., L.N.L., A.S., T.S., A.T., M.C., C.W., J.Z., Y.K., T.J.C., R.-M.M., C.-W.Q., C.L., Y.-W.H., M.A.S.P., I.-C.B.-C., R.G., S.K., A.S.L.G., D.R., Y.W., A.Ay., A.G.S.F., T.C., C.M. contributed to the conceptual discussions, refinement of ideas, and revision of the manuscript.

C.M. conceptualized the main framework, prepared the initial draft, facilitated the community review process, and coordinated the iterative revisions leading to the final version.

**Competing interest**

Cristiano Matricardi, the corresponding author of this paper, worked as an editor at Nature Communications during the preparation of this manuscript. This paper is a community-driven initiative, developed through extensive consultation with researchers, institutions, and stakeholders over several years. The corresponding author's role was to facilitate and synthesize contributions from the broader scientific community, acting as a convener and editor of collective insights. No editorial privileges or influence related to the journal's publication process have been or will be used in connection with this work.

---

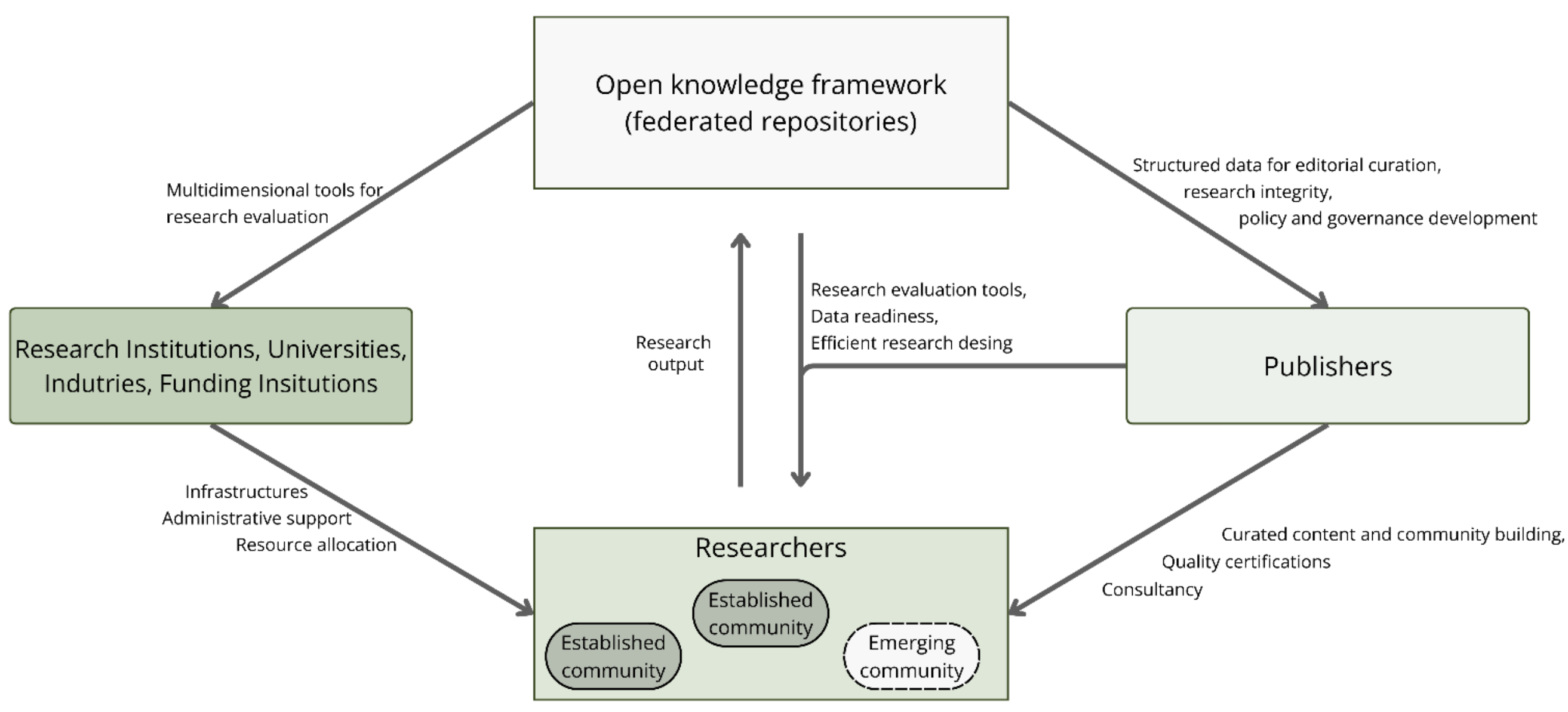

Figure 1. Simplified diagram highlighting the relations between every stakeholder and the conceptualized open knowledge framework. The arrows indicate what a specific group bring to the relative stakeholder.

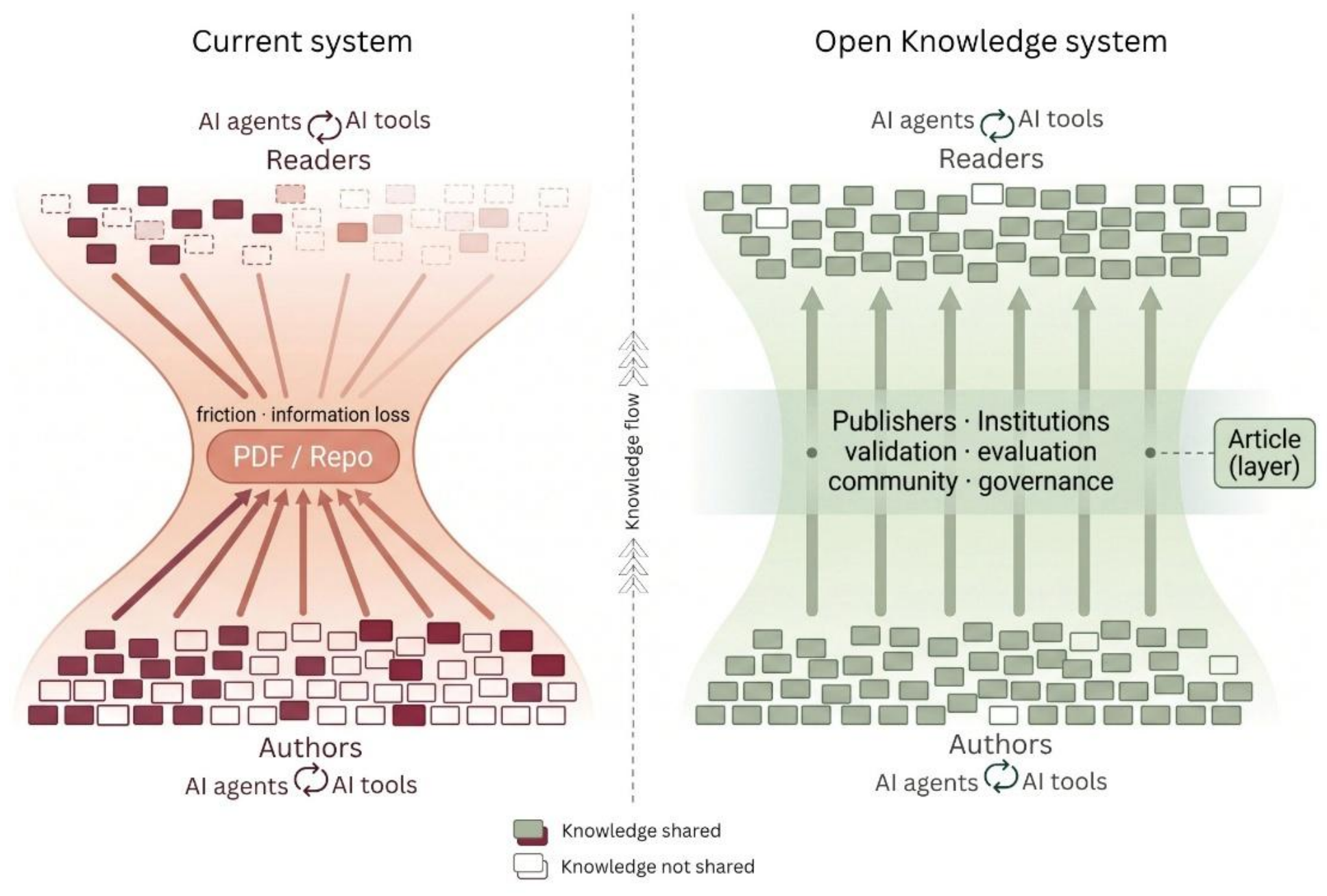

Figure 2. The current and open knowledge systems compared. In the current system (left), most research objects, data, methods, code, negative results, never leave the lab notebook; only a fraction are compressed into a narrative article, creating a bottleneck that generates friction, information loss, and forces readers to reconstruct context. In the open knowledge system (right), structured and interoperable knowledge flows directly from authors to readers and machines through parallel channels, with publishers and institutions acting as a validation and governance layer rather than a gatekeeper. The expansion of shared knowledge objects reflects the core ambition of an open knowledge framework: making the full dimensionality of research legible, verifiable, and reusable across the entire community.

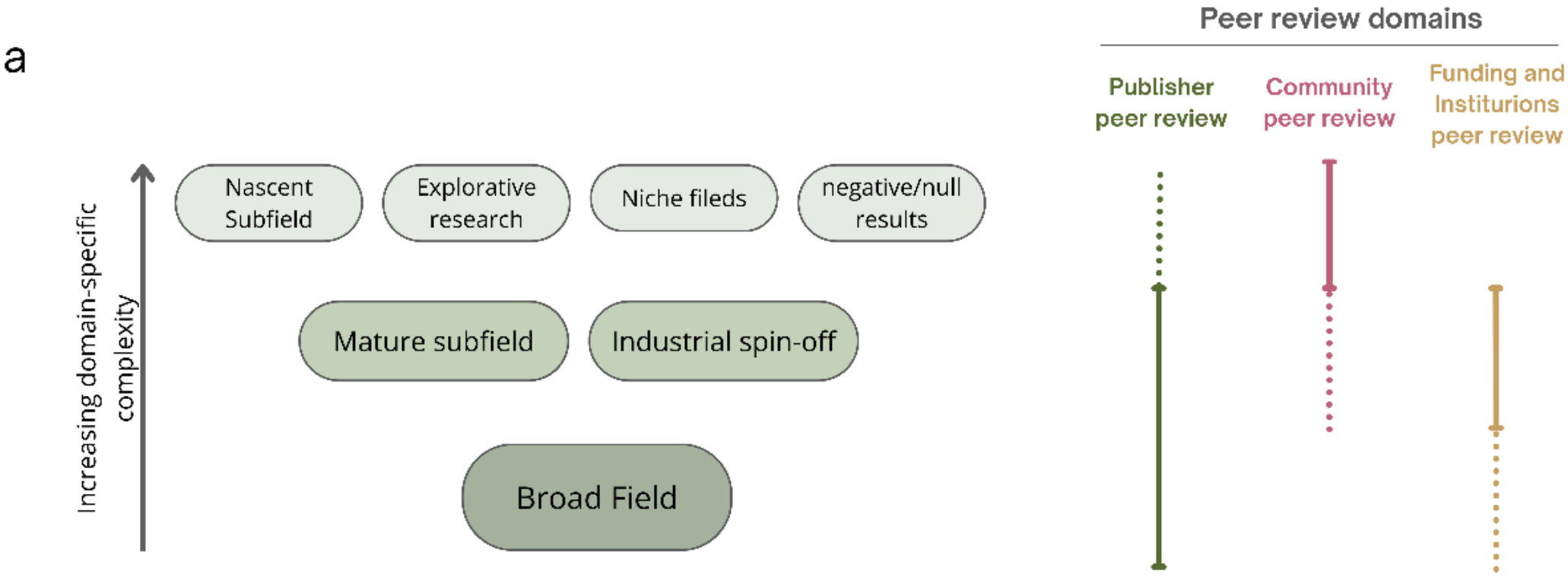

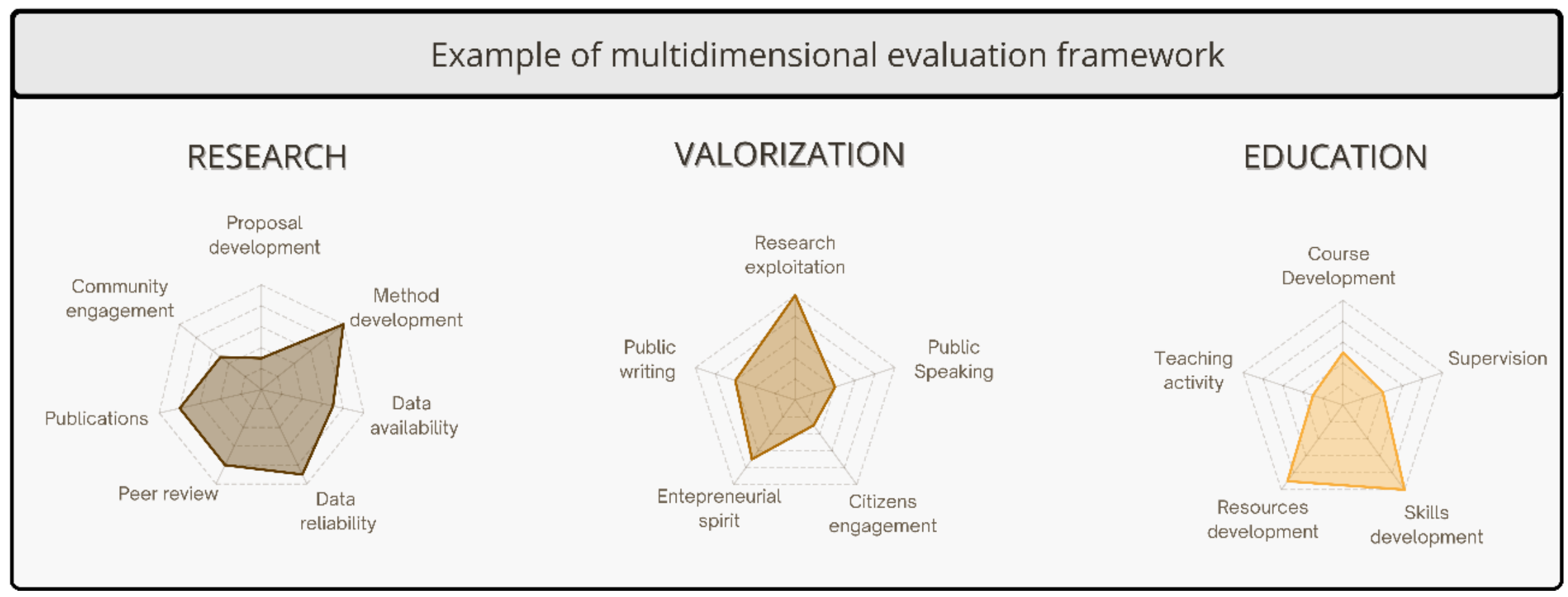

*Figure 3. Schematics illustrations of an evolved per review landscape. a) different domains within the entire scientific ecosystem and the relative adapted peer review strategy, b) outcome of the evaluation process harnessing multidimensional tools for highlighting skills, achievements and attitude. The parameter and evaluation groups have been adapted from the report of the Open and Universal Science Project which identifies indicators and metrics to advance research evaluation*[55]